\documentclass[12pt]{extarticle}

\usepackage[T1]{fontenc}

\usepackage[latin1]{inputenc}
\usepackage{epsfig}
\usepackage[english,french]{babel}
\usepackage{color}
\usepackage{graphicx}

\usepackage{dcolumn}

\usepackage{moreverb}

\usepackage{amsmath,amssymb,amsfonts}

\begin{document}
\numberwithin{equation}{section}
\newcommand{\boxedeqn}[1]{%
  \[\fbox{%
      \addtolength{\linewidth}{-2\fboxsep}%
      \addtolength{\linewidth}{-2\fboxrule}%
      \begin{minipage}{\linewidth}%
      \begin{equation}#1\end{equation}%
      \end{minipage}%
    }\]%
}


\newsavebox{\fmbox}
\newenvironment{fmpage}[1]
     {\begin{lrbox}{\fmbox}\begin{minipage}{#1}}
     {\end{minipage}\end{lrbox}\fbox{\usebox{\fmbox}}}

\begin{flushleft}
\title*{{\LARGE{\textbf{Superintegrability with third order integrals of motion, cubic algebras and supersymmetric quantum mechanics II:Painlev\'e transcendent potentials}}}}
\newline
\newline
Ian Marquette
\newline
D\'epartement de physique et Centre de recherches math\'ematiques,
Universit\'e de Montr\'eal,
\newline
C.P.6128, Succursale Centre-Ville, Montr\'eal, Qu\'ebec H3C 3J7,
Canada
\newline
ian.marquette@umontreal.ca
\newline
\newline
We consider a superintegrable quantum potential in two-dimensional Euclidean space with a second and a third order integral of motion. The potential is written in terms of the fourth Painlev\'e transcendent. We construct for this system a cubic algebra of integrals of motion. The algebra is realized in terms of parafermionic operators and we present Fock type representations which yield the corresponding energy spectra. We also discuss this potential from the point of view of higher order supersymmetric quantum mechanics and obtain ground state wave functions.
\newline
\newline
\section{Introduction}
Over the years many articles have been devoted to superintegrable systems with second order integrals of motion [1-12]. 
Integrable and superintegrable systems with third order integrals have also been studied, albeit to a lesser degree [13,14,15,16,17,18,19]. This article is the second in a series [18] devoted to superintegrable systems in quantum mechanics in two-dimensional Euclidean space $E_{2}$. All classical and quantum potentials with one second and one third order integral of motion that separate in Cartesian coordinates in the two-dimensional Euclidean space were found by S.Gravel [16]. There are 21 quantum potentials and 8 classical ones. The systems investigated were of the form
\newline
\begin{equation}
H=\frac{P_{x}^{2}}{2}+\frac{P_{y}^{2}}{2}+g_{1}(x)+g_{2}(y) \quad ,
\end{equation}
\begin{equation}
A=\frac{P_{x}^{2}}{2}-\frac{P_{y}^{2}}{2}+g_{1}(x)-g_{2}(y)\quad ,
\end{equation}
\begin{equation}
B=\sum_{i+j+k=3}A_{ijk}\{L_{3}^{i},p_{1}^{j}p_{2}^{k}\}+\{l_{1}(x,y),p_{1}\}+\{l_{2}(x,y),p_{2}\}\quad ,
\end{equation}
\newline
where $\{,\}$ is an anticommutator, $L_{3}=xP_{2}-yP_{1}$ is the angular
momentum. The constants $A_{ijk}$ and functions
V, $l_{1}$ and $l_{2}$ are known [16].
\newline
The quantum case contains very interesting potentials written in term of higher transcendental functions.
The irreducible potentials with rational functions were studied [18]. Polynomial algebras [18-27,29,30,31] and the parafermionic realizations of these algebras were found. The parafermionic realizations made it possible to construct Fock type representations and to obtain the energy spectra. We also studied these potentials from the point of view of the supersymmetric quantum mechanics [32-41].
\newline
Among the 21 types of superintegrable quantum potentials 5 of the irreducible ones are
expressed in terms of Painlev\'e transcendents [42]. Let us present one of the superintegrable potentials of Ref.16 written in terms of the fourth Painlev\'e transcendent $P_{4}(z,\alpha,\beta)$:
\newline
\begin{equation}
g_{1}(x)=\frac{\omega^{2}}{2}x^{2}+\epsilon\frac{\hbar\omega}{2}f^{'}(\sqrt{\frac{\omega}{\hbar}}x)+\frac{\omega\hbar}{2}f^{2}(\sqrt{\frac{\omega}{\hbar}}x)+\omega \sqrt{\hbar \omega}xf(\sqrt{\frac{\omega}{\hbar}}x)+\frac{\hbar\omega}{3}(-\alpha+\epsilon) \quad ,            
\end{equation}
\begin{equation}
g_{2}(x)=\frac{\omega^{2}}{2}y^{2} \quad ,
\end{equation}
where $\epsilon=\pm 1$, $f'=\frac{df}{dz}$, $z=\sqrt{\frac{\omega}{\hbar}}x$
\begin{equation}
f^{''}(z) = \frac{f^{'2}(z)}{2f(z)} + \frac{3}{2}f^{3}(z) + 4zf^{2}(z) + 2(z^{2} -
\alpha)f(z) +  \frac{\beta}{f(z)} \quad ,
\end{equation}
\newline
\begin{equation}
f(z)=P_{4}(z,\alpha,\beta).
\end{equation}
\newline
The six Painlev\'e transcendent functions appear in the theory of nonlinear differential equations. The occurence of Painlev\'e transcendents as superintegrable potentials seems somewhat surprising. It is less so once we remember the relation between the Schrödinger equation and the Korteweg-de Vries equation [43]. Solutions of the KdV include Painlev\'e transcendents. Unidimensional potentials expressed in terms of Painlev\'e transcendents were also obtained in the context of the dressing chains method [44,45,46] and conditionals and higher symmetries [47]. An important aspect of the fourth Painlev\'e transcendent is the existence of particular solutions in terms of rational functions and classical special functions for very specific values of the two parameters $\alpha$ and $\beta$ [48]. 
\newline
All Hamiltonians of Ref.16 are, by construction, the sum of two unidimensional Hamiltonians ($H=H^{x}+H^{y}$). All the quantum potentials with rational function were related to supersymmetric quantum mechanics [19]. Higher order SUSYQM and shape invariance have been investigated [49,50,51,52,53,54,55]. In the case of the potential given by Eq.(1.4) and (1.5) the Hamiltonian $H^{y}$ is the well known harmonic oscillator. The Hamiltonian $H^{x}$ the corresponding Schrödinger equation has been obtained as a special case of third order shape invariance and solved [51]. 
\newline
This article is organized in the following way. In Section 2 we construct the Fock type representations for the superintegrable potential given by the Eq(1.1) by the means of realizations of cubic algebras in terms of a parafermionic algebra. In the Section 3 we will recall some aspects of third order shape invariance that are related to the potential given by Eq.(1.4) and (1.5) with $\epsilon=-1$. We will also treat the case with $\epsilon=1$. We will relate these results to those obtained using the approach involving the cubic algebra. In Section 4 we will consider special cases and apply results of Section 2 and Section 3.
\newline
\section{Cubic and parafermionic algebras}
We consider a quantum superintegrable Hamiltonian in E2 involving the fourth Painlev\'e trascendent. We have two cases $\epsilon$=1 and $\epsilon$=-1 (with $\omega>0$)
\newline
\newline
\begin{equation}
H=\frac{P_{x}^{2}}{2}  + \frac{P_{y}^{2}}{2}
+\frac{\omega^{2}}{2}y^{2}+g_{1}(x) \quad ,
\end{equation}
with $g_{1}(x)$ given in (1.4). This Hamiltonian has two integrals of motion. The one of the second order is given by Eq.(1.2) and Eq.(1.4). The third order one is given by the following equation:
\newline
\begin{equation}
B=
\frac{1}{2}\{L,P_{x}^{2}\}+\frac{1}{2}\{\frac{\omega^{2}}{2}x^{2}y
-3yg_{1}(x),P_{x}\}-\frac{1}{w^{2}}\{\frac{\hbar^{2}}{4}g_{1xxx}(x)+(\frac{\omega^{2}}{2}x^{2}-3g_{1}(x))g_{1x}(x),P_{y}\},
\end{equation}
\newline
\newline
where $L=xP_{y}-yP{x}$. 
\newline
The operators A and B generate the following cubic algebra
\newline
\begin{equation}
 [A,B]\equiv C \quad [A,C]=16\omega^{2}\hbar^{2}B
\end{equation}
\[  [B,C]=-2\hbar^{2}A^{3}-6\hbar^{2}HA^{2}+8\hbar^{2}H^{3}   \]
\[ + \frac{\omega^{2}\hbar^{4}}{3}(4\alpha^{2}-20-6\beta-8\epsilon\alpha)A-8\omega^{2}\hbar^{4}H    \]
\[ +\frac{\hbar^{5}\omega^{3}}{27}(-8\alpha^{3}-24\alpha-36\alpha\beta+24\epsilon\alpha^{2}+8\epsilon+36\epsilon\beta)\quad .   \]
\newline
The Casimir operator can be written as a polynomial in the Hamiltonian
\newline
\begin{equation}
K=-16\hbar^{2}H^{4}+\frac{4\hbar^{4}\omega^{2}}{3}(4\alpha^{2}-8\alpha+4-\alpha\beta )H^{2}
\end{equation}
\[-\frac{4\hbar^{5}\omega^{3}}{27}(8\alpha^{3}-24\epsilon\alpha^{2}+24\alpha+36\alpha\beta-8\epsilon-36\epsilon\beta)H             \]
\[-\frac{4\hbar^{6}\omega^{4}}{3}(4\alpha-8\epsilon\alpha-8-6\beta) \quad .          \]
\newline
Realizations of cubic algebras in terms of parafermionic algebras have been discussed in our previous article [19]. Our potential belong to the Case 2 of Ref.19. The cubic algebra has the form:
\newline
\begin{equation}
[A,B]=C, \quad [A,C]= \delta B,\quad [B,C]=\mu A^{3} + \nu A^{2} + \xi A + \zeta \quad ,
\end{equation}
\newline
where 
\begin{equation}
\mu=\mu_{0},\quad \nu=\nu_{0}+\nu_{1}H,\quad \xi=\xi_{0}+\xi_{1}H+\xi_{2}H^{2}
\end{equation}
\[\zeta=\zeta_{0}+\zeta_{1}H+\zeta_{2}H^{2}+\zeta_{3}H^{3},\quad \delta=\delta_{0}+\delta_{1}H \quad .\]
\newline
This algebra has been realized in terms of a deformed oscillator algebra of the form
\begin{equation}
[N,b^{t}]=b^{t} ,\quad [N,b]=-b ,\quad b^{t}b=\Phi(N) ,\quad
bb^{t}=\Phi(N+1) \quad .
\end{equation}
The structure function $\Phi(N)$ is given by
\newline
\begin{equation}
\Phi(N) = (\frac{K}{-4\delta}-\frac{\zeta}{4\sqrt{\delta}})+(-\frac{\xi}{4}+\frac{\zeta}{2\sqrt{\delta}}+\frac{\nu\sqrt{\delta}}{12})(N+u)
\end{equation}
\[+(\frac{-\nu\sqrt{\delta}}{4}+\frac{\xi}{4}+\frac{\mu\delta}{8})(N+u)^{2}+(\frac{\nu\sqrt{\delta}}{6}-\frac{\mu \delta}{4})(N+u)^{3}+(\frac{\mu\delta}{8})(N+u)^{4} \quad . \]
\newline
We can use Eq.(2.4) to rewrite the structure function in terms of the Hamiltonian.
\subsection{Case $\epsilon = 1$.}
From the general formula we obtain for our particular case the following structure function for $\epsilon =1$
\newline
\begin{equation}
\Phi(x)=-4\omega^{2}\hbar^{4}(x+u-(\frac{E}{2\hbar\omega}+\frac{1}{2}))(x+u-(\frac{-E}{2\hbar\omega}+\frac{5}{6}-\frac{\alpha}{3}))
\end{equation}
\[ (x+u-(\frac{-E}{2\hbar\omega}+\frac{1}{6}(\alpha+2-3i\sqrt{\frac{\beta}{2}})))  (x+u-(\frac{-E}{2\hbar\omega}+\frac{1}{6}(\alpha+2+3i\sqrt{\frac{\beta}{2}})))    \quad .     \]
\newline
To obtain unitary representations [25,26] we should impose three constraints given by 
\newline
\begin{equation}
\Phi(p+1,u_{i},k)=0, \quad \Phi(0,u,k)=0,\quad \phi(x)>0, \quad \forall \quad x>0 \quad .
\end{equation}
\newline
We have to distinguish the two cases $\beta < 0$ and $\beta > 0$. For $\beta < 0$
we get four possible values for u with $\Phi(0,u,k)=0$
\newline
\begin{equation}
u_{1}=\frac{-E}{2\hbar\omega}+\frac{5}{6}-\frac{\alpha}{3},\quad u_{2}=\frac{-E}{2\hbar\omega}+\frac{1}{6}(\alpha+2+3\sqrt{\frac{-\beta}{2}})
\end{equation}
\[ u_{3}=\frac{-E}{2\hbar\omega}+\frac{1}{6}(\alpha+2-3\sqrt{\frac{-\beta}{2}}),\quad u_{4}=\frac{E}{2\hbar\omega}+\frac{1}{2}   \quad .           \]
\newline
We insert all these solutions for u and apply the constraint $\Phi(p+1,u_{i},k)=0$, with i=1,2,3,4 to find the energy spectrum.
\newline
\textbf{Case 1}
\newline
\begin{equation}
\Phi(x)=4\hbar^{4}\omega^{2}x(p+1-x)(x+\frac{1}{2}-\frac{\alpha}{2}-\sqrt{\frac{-\beta}{8}})(x+\frac{1}{2}-\frac{\alpha}{2}+\sqrt{\frac{-\beta}{8}})
\end{equation}
\begin{equation}
E=\hbar\omega(p+\frac{4}{3}-\frac{\alpha}{3}).
\end{equation}
\newline
\textbf{Case 2}
\newline
\begin{equation}
\Phi(x)=4\hbar^{4}\omega^{2}x(p+1-x)(x+\sqrt{\frac{-\beta}{2}})(x-\frac{1}{2}+\frac{\alpha}{2}+\sqrt{\frac{-\beta}{8}})
\end{equation}
\begin{equation}
E=\hbar\omega(p+\frac{5}{6}+\frac{\alpha}{6}+\sqrt{\frac{-\beta}{8}}).
\end{equation}
\newline
\textbf{Case 3}
\newline
\begin{equation}
\Phi(x)=4\hbar^{4}\omega^{2}x(p+1-x)(x-\sqrt{\frac{-\beta}{2}})(x-\frac{1}{2}+\frac{\alpha}{2}-\sqrt{\frac{-\beta}{8}}),
\end{equation}
\begin{equation}
E=\hbar\omega(p+\frac{5}{6}+\frac{\alpha}{6}-\sqrt{\frac{-\beta}{8}}).
\end{equation}
\newline
\textbf{Case 4}
\newline
\newline
We get three solutions for this case with negative energy
\newline
\begin{equation}
\Phi(x)=4\hbar^{4}\omega^{2}x(p+1-x)((p+\frac{1}{2})+\frac{\alpha}{2}+\sqrt{\frac{-\beta}{8}}-x)((p+\frac{1}{2})+\frac{\alpha}{2}-\sqrt{\frac{-\beta}{8}}-x),
\end{equation}
\begin{equation}
E=-\hbar\omega(p+\frac{2}{3}+\frac{\alpha}{3}),
\end{equation}
\newline
\begin{equation}
\Phi(x)=4\hbar^{4}\omega^{2}x(p+1-x)((p+\frac{3}{2})-\frac{\alpha}{2}-\sqrt{\frac{-\beta}{8}}-x)((p+1)-\sqrt{\frac{-\beta}{2}}-x),
\end{equation}
\begin{equation}
E=-\hbar\omega(p+\frac{7}{6}-\frac{\alpha}{6}-\sqrt{\frac{-\beta}{8}}),
\end{equation}
\newline
\begin{equation}
\Phi(x)=4\hbar^{4}\omega^{2}x(p+1-x)((p+\frac{3}{2})-\frac{\alpha}{2}+\sqrt{\frac{-\beta}{8}}-x)((p+1)+\sqrt{\frac{-\beta}{2}}-x),
\end{equation}
\begin{equation}
E=-\hbar\omega(p+\frac{7}{6}-\frac{\alpha}{6}+\sqrt{\frac{-\beta}{8}}).
\end{equation}
\newline
\newline
To obtain unitary representions we should also impose $\phi(x)$ to be a real function and $\phi(x)>0$ for $x>0$. The constraints do not allow all values for $\alpha$ and $\beta$ so it may happen that only some of the states are physically meaningful. We can have 1, 2 or 3 infinite sequences of energies that correspond to each unitary representation. 
\newline
For $\beta > 0$ we have two solutions
\begin{equation}
\Phi(x)=4\hbar^{4}\omega^{2}x(p+1-x)(x^{2}+(1-\alpha)x-\frac{\beta}{8}+\frac{\alpha^{2}}{4}-\frac{\alpha}{2}+\frac{1}{4}),
\end{equation}
\begin{equation}
E=\hbar\omega(p+\frac{4}{3}-\frac{\alpha}{3}),
\end{equation}
\begin{equation}
\Phi(x)=4\hbar^{4}\omega^{2}x(p+1-x)(x^{2}-(1+\alpha+2p)x+p^{2}+\alpha p +p-\frac{\beta}{8}+\frac{\alpha^{2}}{4}+\frac{\alpha}{2}+\frac{1}{4}),
\end{equation}
\begin{equation}
E=-\hbar\omega(p+\frac{2}{3}+\frac{\alpha}{3}).
\end{equation}
\newline
\newline
\subsection{Case $\epsilon = -1$.}
For the case $\epsilon$=-1 we obtain the following expression for the structure function
\newline
\begin{equation}
\Phi(x)=-4\omega^{2}\hbar^{4}(x+u-(\frac{E}{2\hbar\omega}+\frac{1}{2}))(x+u-(\frac{-E}{2\hbar\omega}+\frac{1}{6}-\frac{\alpha}{3}))
\end{equation}
\[ (x+u-(\frac{-E}{2\hbar\omega}+\frac{1}{6}(\alpha+4-3i\sqrt{\frac{\beta}{2}})))(x+u-(\frac{-E}{2\hbar\omega}+\frac{1}{6}(\alpha+4+3i\sqrt{\frac{\beta}{2}})))   \quad .         \]
\newline
Four cases occur for $\beta < 0$
\newline
\begin{equation}
u_{1}=\frac{-E}{2\hbar\omega}+\frac{1}{6}-\frac{\alpha}{3},\quad u_{2}=\frac{-E}{2\hbar\omega}+\frac{1}{6}(\alpha+4+3\sqrt{\frac{-\beta}{2}})
\end{equation}
\[ u_{3}=\frac{-E}{2\hbar\omega}+\frac{1}{6}(\alpha+4-3\sqrt{\frac{-\beta}{2}}) \quad u_{4}=\frac{E}{2\hbar\omega}+\frac{1}{2}  \quad .              \]
\newline
\textbf{Case 1}
\newline
\begin{equation}
\Phi(x)=4\hbar^{4}\omega^{2}x(p+1-x)(x-\frac{1}{2}-\frac{\alpha}{2}-\sqrt{\frac{-\beta}{8}})(x-\frac{1}{2}-\frac{\alpha}{2}+\sqrt{\frac{-\beta}{8}}),
\end{equation}
\begin{equation}
E=\hbar\omega(p+\frac{2}{3}-\frac{\alpha}{3}).
\end{equation}
\newline
\textbf{Case 2}
\newline
\begin{equation}
\Phi(x)=4\hbar^{4}\omega^{2}x(p+1-x)(x+\sqrt{\frac{-\beta}{2}})(x+\frac{1}{2}+\frac{\alpha}{2}+\sqrt{\frac{-\beta}{8}}),
\end{equation}
\begin{equation}
E=\hbar\omega(p+\frac{7}{6}+\frac{\alpha}{6}+\sqrt{\frac{-\beta}{8}}).
\end{equation}
\newline
\textbf{Case 3}
\newline
\begin{equation}
\Phi(x)=4\hbar^{4}\omega^{2}x(p+1-x)(x-\sqrt{\frac{-\beta}{2}})(x+\frac{1}{2}+\frac{\alpha}{2}-\sqrt{\frac{-\beta}{8}}),
\end{equation}
\begin{equation}
E=\hbar\omega(p+\frac{7}{6}+\frac{\alpha}{6}-\sqrt{\frac{-\beta}{8}}).
\end{equation}
\newline
\textbf{Case 4}
\newline
We get three solutions for this case with negative energy
\newline
\begin{equation}
\Phi(x)=4\hbar^{4}\omega^{2}x(p+1-x)((p+\frac{3}{2})+\frac{\alpha}{2}-\sqrt{\frac{-\beta}{8}}-x)((p+\frac{3}{2})+\frac{\alpha}{2}+\sqrt{\frac{-\beta}{8}}-x),
\end{equation}
\begin{equation}
E=-\hbar\omega(p+\frac{4}{3}+\frac{\alpha}{3}),
\end{equation}
\newline
\begin{equation}
\Phi(x)=4\hbar^{4}\omega^{2}x(p+1-x)((p+\frac{1}{2})-\frac{\alpha}{2}-\sqrt{\frac{-\beta}{8}}-x)((p+1)-\sqrt{\frac{-\beta}{2}}-x),
\end{equation}
\begin{equation}
E=-\hbar\omega(p+\frac{5}{6}-\frac{\alpha}{6}-\sqrt{\frac{-\beta}{8}}),
\end{equation}
\newline
\begin{equation}
\Phi(x)=4\hbar^{4}\omega^{2}x(p+1-x)((p+\frac{1}{2})-\frac{\alpha}{2}-\sqrt{\frac{-\beta}{8}}-x)((p+1)+\sqrt{\frac{-\beta}{2}}-x),
\end{equation}
\begin{equation}
E=-\hbar\omega(p+\frac{5}{6}-\frac{\alpha}{6}+\sqrt{\frac{-\beta}{8}}).
\end{equation}
\newline
One interesting aspect of this potential is that we can have three, two or one series of equidistant energy levels. 
\newline 
For $\beta > 0$ we get the following solution
\newline
\begin{equation}
\Phi(x)=4\hbar^{4}\omega^{2}x(p+1-x)(x^{2}-(1+\alpha)x-\frac{\beta}{8}+\frac{\alpha^{2}}{4}+\frac{\alpha}{2}+\frac{1}{4}),
\end{equation}
\begin{equation}
E=\hbar\omega(p+\frac{2}{3}-\frac{\alpha}{3}),
\end{equation}
\begin{equation}
\Phi(x)=4\hbar^{4}\omega^{2}x(p+1-x)(x^{2}-(3+\alpha+2p)x+p^{2}+\alpha p +3p-\frac{\beta}{8}+\frac{\alpha^{2}}{4}+\frac{\alpha}{2}+\frac{9}{4}),
\end{equation}
\begin{equation}
E=-\hbar\omega(p+\frac{4}{3}+\frac{\alpha}{3}).
\end{equation}
\newline
\section{Third order shape invariance and superintegrable systems}
The concept of higher-derivative supersymmetric quantum mechanics (HSQM) was introduced by A.A.Andrianov, M.V.Ioffe and V.P.Spiridonov [49]. HSQM is characterized by polynomial relations between supercharges and the Hamiltonian. Second order derivative supersymmetry was investigated in the Ref.50. We will present in this section results not given in Ref.51 but directly related to the potential given by Eq.(1.1) with $\epsilon=1$. Let us recall some aspects of the particular case of third order shape invariance related to the potential with $\epsilon=-1$ obtained in the Ref.51. In SUSYQM two superpartners are isospectral or almost isospectral and if we know the spectrum and the eigenfunctions of one superpartner we can obtain the spectrum and the eigenfunctions of the other superpartner. A special case occurs when the two superpartners $V_{1}(x,a_{0})$ and $V_{2}(x,a_{0})$ satisfy the relation $V_{2}(x,a_{1})=V_ {1}(x,a_{0})+R(a_{1})$ where $a_{1}=f(a_{0})$ and $R(a_{1})$ do not depend on x. In this special case we can find directly the energy and the eigenfunctions. The superpartners are called shape invariant potentials (SIP). We consider the following particular case of shape invariance 
\begin{equation}
H_{1}a^{\dagger}=a^{\dagger}(H_{1}+2\lambda) \quad ,
\end{equation}
where $a^{\dagger}$ and $a$ are third order operators. This particular case of shape invariance can be constructed from a first order and second order supersymmetry given by the following interwining relations
\newline
\begin{equation}
H_{1}q^{\dagger}=q^{\dagger}(H_{2}+2\lambda),\quad H_{1}M^{\dagger}=M^{\dagger}H_{2}\quad ,
\end{equation}
where
\newline
\begin{equation}
H_{i}=P_{x}^{2}+V_{i}(x)\quad ,
\end{equation}
\begin{equation}
q^{\dagger}=\partial + W(x),\quad q=-\partial +W(x)\quad ,
\end{equation}
\begin{equation}
M^{\dagger}=\partial^{2}-2h(x)\partial+b(x),\quad M=\partial^{2}+2h(x)\partial+b(x)\quad .
\end{equation}
The key element in obtaining the equivalence between Eq.(3.1) and Eq.(3.2) is to define the following third order operators $a$ and $a^{\dagger}$ written as products of first order and second order supercharges
\begin{equation}
a^{\dagger}=q^{\dagger}M,\quad a=M^{\dagger}q \quad .
\end{equation}
The third order shape invariance of the form given by Eq.(3.1) can be investigated using Eq.(3.2). The two interwining relations of Eq.(3.2) give respectively the following relations
\newline
\begin{equation}
V_{1}=W'(x)+W^{2}(x),\quad V_{2}=-W'(x)+W^{2}(x)-2\lambda
\end{equation}
and
\begin{equation}
V_{1,2}=\mp 2h'(x)+h^{2}(x)+\frac{h''(x)}{2h(x)}-\frac{h^{'2}(x)}{4h^{2}(x)}-\frac{d}{4h^{2}(x)}+\gamma \quad ,
\end{equation}
\begin{equation}
b(x)=-h^{'}(x)+h^{2}(x)-\frac{h^{''}(x)}{2h(x)}+\frac{h^{'2}(x)}{4h^{2}(x)}+\frac{d}{4h^{2}(x)} \quad .
\end{equation}
Eq.(3.7), (3.8) and (3.9) impose that the potential $V_{1}$ should have the form
\newline
\begin{equation}
V_{1}=-2h'(x)+4h^{2}(x)+4\lambda x h(x)+\lambda^{2}x^{2}-\lambda \quad ,
\end{equation}
with
\begin{equation}
h''(x)=\frac{h^{'2}(x)}{2h(x)}+6h^{3}(x)+8\lambda xh^{2}(x)+2(\lambda^{2}x^{2}-(\lambda+\gamma))h(x)+\frac{d}{2h(x)},
\end{equation}
\begin{equation}
W(x)=W_{3}(x)=-2h(x)-\lambda x \quad .
\end{equation}
As in the case of first order supersymmetry we can define
\newline
\begin{equation}
H = \begin{pmatrix}H_{1} & 0 \\ 0 & H_{2}\end{pmatrix}\quad Q = \begin{pmatrix}0 & 0 \\ M & 0\end{pmatrix}\quad Q^{\dagger} = \begin{pmatrix} 0 & M^{\dagger} \\ 0 & 0\end{pmatrix}\quad .
\end{equation}
\newline
We get the following SUSY-algebra
\newline
\begin{equation}
[H,Q]=[H,Q^{\dagger}]=0, \quad  \{Q,Q\}=\{Q^{\dagger},Q^{\dagger}\}=0, \quad  \{Q,Q^{\dagger}\}=(H-\gamma)^{2}+d \quad .
\end{equation}
\newline
Eq.(3.11) can be transformed into the equation for the fourth Painlev\'e transcendent (1.4) by the following transformations 
\begin{equation}
h(x)=\frac{1}{2}\sqrt{\lambda}f(z),\quad z=\sqrt{\lambda}x,\quad \alpha=1+\frac{\gamma}{\lambda},\quad \beta=\frac{2d}{\lambda^{2}},\quad \lambda=\frac{\omega}{\hbar},
\end{equation}
and we obtain
\begin{equation}
\tilde{V_{1}}=\frac{\hbar^{2}}{2}V_{1}=\frac{\omega^{2}}{2}x^{2}-\frac{\omega\hbar}{2}f'(\sqrt{\frac{\omega}{\hbar}}x)+\frac{\omega\hbar}{2}f^{2}(\sqrt{\frac{\omega}{\hbar}}x)+\omega\sqrt{\omega\hbar}xf(\sqrt{\frac{\omega}{\hbar}}x)-\omega\hbar .
\end{equation}
\newline
$\tilde{V_{1}}(x)$ is the x part of the potential in (1.1) and coincides with $g_{1}(x)$ in Eq.(1.4) up to a constant.
A particular case of third order shape invariance called \og reducible\fg was considered in Ref.51 by imposing further conditions. These conditions are d $\leq$ 0 and the existence of real functions $W_{1}$ and $W_{2}$ such that
\begin{equation}
M^{\dagger}=(\partial+W_{1}(x))(\partial+W_{2}(x)), \quad W_{1,2}=-h(x)\pm \frac{h^{'}(x)-\sqrt{-d}}{2h(x)}\quad ,
\end{equation}
(reducible means that $M^{\dagger}$ factorizes into product of two first order operators with real functions). The spectrum was obtained for cases where normalizable zero modes of the annihilation operator exist. Zero modes of the annihilation operator satisfy
\begin{equation}
a\psi_{k}^{(0)}=0 \quad ,
\end{equation}
(we use the terminology of HSQM where \og zero mode \fg refers to Eq.(3.18) so that zero modes may not have energy $E_{0}=0$). The energies of zero modes were obtained by imposing the vanishing of the norm of $a\psi_{k}^{(0)}$ which involves the average of the operator product $a^{\dagger}a$.
\newline
\begin{equation}
a^{\dagger}a=q^{\dagger}MM^{\dagger}q=q^{\dagger}((H_{2}-\gamma)^{2}+d)q=H_{1}((H_{1}-\gamma-2\lambda)^{2}+d)\quad .
\end{equation}
The energies of the zero modes are
\newline
\begin{equation}
E_{1}^{(0)}=0,\quad E_{2}^{(0)}=\gamma+2\lambda+\sqrt{-d},\quad E_{3}^{(0)}=\gamma+2\lambda-\sqrt{-d}\quad .
\end{equation}
The corresponding eigenfunctions $\psi_{k}^{(0)}$ can be calculated explicitly and are
\begin{equation}
\psi_{1}^{0}(x)=e^{\int^{x}W_{3}(x')dx'}\quad ,
\end{equation}
\begin{equation}
\psi_{2}^{0}(x)=(W_{2}(x)-W_{3}(x))e^{-\int^{x}W_{2}(x')dx'}\quad ,
\end{equation}
\begin{equation}
\psi_{3}^{0}(x)=(2\sqrt{-d}+(W_{2}(x)-W_{3}(x))(W_{1}(x)+W_{2}(x)))e^{-\int^{x}W_{1}(x')dx'}\quad .
\end{equation}
\newline
The creation operator can also have zero modes $\phi_{k}^{(0)}$ which correspond to a possible truncation of the sequence of excited levels. They were obtained by considering the following product
\begin{equation}
aa^{\dagger}=(H_{1}+2\lambda)((H_{1}-\gamma)^{2}+d) \quad .
\end{equation}
The energies of the zero modes are
\begin{equation}
E_{1}^{(0)}=\gamma-\sqrt{-d},\quad E_{2}^{(0)}=\gamma+\sqrt{-d},\quad E_{3}^{(0)}=-2\lambda \quad ,
\end{equation}
with the corresponding eigenfunctions
\begin{equation}
\phi_{1}^{0}(x)=e^{\int^{x}W_{1}(x')dx'}\quad ,
\end{equation}
\begin{equation}
\phi_{2}^{0}(x)=(W_{1}(x)+W_{2}(x))e^{\int^{x}W_{2}(x')dx'}\quad ,
\end{equation}
\begin{equation}
\phi_{3}^{0}(x)=(\gamma+2\lambda+\sqrt{-d}+(W_{1}(x)+W_{2}(x))(W_{2}(x)-W_{3}(x)))e^{-\int^{x}W_{3}(x')dx'}\quad .
\end{equation}
For non singular potentials it is not possible to have the negative energy $E_{3}^{(0)}$ and the total number of zero modes of the annihilation and creation operator cannot be more than three because of the asymptotics of the eigenfunctions. We can have three, two or one infinite sequence of levels. These results coincide with those obtained as from the analysis of Fock type representations of the cubic algebra of the superintegrable potential. When we apply the creation operator $a^{\dagger}$ on zero modes we create eigenfunctions with $2\lambda$ more energy. These energies are corroborated (when we add a harmonic oscillator in the y direction) by those obtained using the cubic algebra and given by Eq.(2.31), Eq.(2.33) and Eq.(2.35).
\newline 
When a potential allowes only one infinite sequence of energies, this potential may also allow a singlet state or a doublet of states
\newline
\begin{equation}
a^{+}\psi(x)=a^{-}\psi(x)=0,\quad (a^{+})^{2}\psi(x)=a^{-}\psi(x)=0\quad .
\end{equation}
\newline
From an algebraic point of view these states correspond to trivial irreducible representations. Such a case was discussed in Ref. [19] for the potential $V=\hbar^{2}(\frac{x^{2}+y^{2}}{8a^{4}} + \frac{1}{(x-a)^{2}}+\frac{1}{(x+a)^{2}})$. This potential is a special case of the potential given by the Eq(1.1). The observed singlet state can now be naturally understood as a phenomenon of third order shape invariance. 
\newline
All the results we presented apply to our potential for $\epsilon=-1$. We will present here the results that will be applicable to the case $\epsilon=1$. We follow the same approach as in Ref.51 and we consider the following potential 
\newline
\begin{equation}
V_{2}=2h'(x)+4h^{2}(x)+4\lambda x h(x)+\lambda^{2}x^{2}-\lambda \quad .
\end{equation}
The Eq.(3.9), Eq.(3.11) and Eq.(3.12) remain the same. We can define as for $V_{1}$ in Eq.(3.16) a potential $\tilde{V_{2}}$ using the transformations of Eq.(3.15). The Hamiltonian $H_{2}$ of the form given by Eq.(3.3) with the potential given by Eq.(3.30) satisfies 
\newline
\begin{equation}
H_{2}a^{\dagger}=a^{\dagger}(H_{2}+2\lambda) \quad .
\end{equation}
when we postulate
\begin{equation}
a^{\dagger}=Mq^{\dagger},\quad a=qM^{\dagger} \quad .
\end{equation}
We have the following product
\begin{equation}
a^{\dagger}a=H_{2}((H_{2}-\gamma)^{2}+d),\quad aa^{\dagger}=(H_{2}+2\lambda)((H_{2}+2\lambda-\gamma)^{2}+d) \quad .
\end{equation}
The energies of the zero modes of the creation and annihilation operator are obtained by imposing the vanishing of their norm. This involves the average of the operator product $a^{\dagger}a$ and $aa^{\dagger}$ given by Eq.(3.33). The eigenfunctions of the zero modes are obtained directly by solving $a\psi_{k}^{(0)}=0$ and $a^{\dagger}\phi_{k}^{(0)}=0$. The energies of zero modes of the annihilation operator are 
\newline
\begin{equation}
E_{1}^{(0)}=0,\quad E_{2}^{(0)}=\gamma-\sqrt{-d},\quad E_{3}^{(0)}=\gamma+\sqrt{-d} \quad ,
\end{equation}
with the corresponding eigenfunctions
\begin{equation}
\psi_{1}^{0}(x)=(\gamma-\sqrt{-d}+(W_{1}(x)+W_{2}(x))(W_{1}(x)-W_{3}(x)))e^{\int^{x}W_{3}(x')dx'}\quad ,
\end{equation}
\begin{equation}
\psi_{2}^{0}(x)=(W_{1}(x)+W_{2}(x))e^{-\int^{x}W_{1}(x')dx'}\quad ,
\end{equation}
\begin{equation}
\psi_{3}^{0}(x)=e^{-\int^{x}W_{2}(x')dx'} \quad .
\end{equation}
The energies of zero modes of the creation operator are
\newline
\begin{equation}
E_{1}^{(0)}=-2\lambda,\quad E_{2}^{(0)}=\gamma-2\lambda-\sqrt{-d},\quad E_{3}^{(0)}=\gamma-2\lambda+\sqrt{-d} \quad ,
\end{equation}
with the corresponding eigenfunctions
\begin{equation}
\phi_{1}^{0}(x)=e^{-\int^{x}W_{3}(x')dx'} \quad ,
\end{equation}
\begin{equation}
\phi_{2}^{0}(x)=(W_{1}(x)-W_{3}(x))e^{\int^{x}W_{1}(x')dx'} \quad ,
\end{equation}
\begin{equation}
\phi_{3}^{0}(x)=(-2\sqrt{-d}+(W_{1}(x)-W_{3}(x))(W_{1}(x)+W_{2}(x)))e^{\int^{x}W_{2}(x')dx'} \quad .
\end{equation}
Again the total number of zero modes of the annihilation and creation operator cannot be more than three because of the asymptotics of the eigenfunctions. We can have three, two or one infinite sequence of levels. When we apply the creation operator $a^{\dagger}$ on zero modes we create eigenfunctions with $2\lambda$ more energy. These energies are corroborated (when we add a harmonic oscillator in the y direction) by those obtained by the cubic algebra and given by Eq.(2.13), Eq.(2.15) and Eq.(2.17). When a potential possess only one infinite sequence of energies, this potential may also possess a singlet state
or a doublet states.
\newline
\newline
We will discuss the irreducible case that appears when $d > 0$. For $V_{1}(x)$ we get
\newline
\begin{equation}
E_{1}^{(0)}=0,\quad \psi_{0}^{0}(x)=e^{\int^{x}W_{3}(x')dx'}.
\end{equation}
For $V_{2}(x)$ we get
\newline
\begin{equation}
E_{1}^{(0)}=0,\quad \psi_{0}^{0}(x)=(\gamma-\sqrt{d}+(W_{1}(x)+W_{2}(x))(W_{1}(x)-W_{3}(x)))e^{\int^{x}W_{3}(x')dx'}.
\end{equation}
and $W_{1}$ and $W_{2}$ are not real functions.
\newline
\section{Special cases}
The fourth Painlev\'e transcendent satisfying Eq.(1.6) depends on two parameters and special solutions in terms of rational or classical special functions exist [48]. In this section, we will give the unitary representations, the corresponding energy spectra and the eigenfunctions for some special cases.
\newline
\subsection{Case  $ \alpha = 5$, $\beta = -8$, $f(z)=\frac{4z(2z^{2} - 1)(2z^{2}+ 3)}{(2z^{2}+ 1)(4z^{4}+3)}$ and $\epsilon = 1$.}
We have with Eq.(1.4) and (1.5)
\newline
\begin{equation}
V(x,y)=\frac{\omega^{2}}{2}(x^{2}+y^{2})-\frac{8\hbar^{3}\omega}{(2\omega x^{2}+\hbar)^{2}}+\frac{4\hbar^{2}\omega}{(2\omega x^{2}+\hbar)}+\frac{2\hbar\omega}{3}.
\end{equation}
\newline
From the cubic algebra we get two unitary representations. The first unitary representation is given by Eq.(2.14) with the corresponding energy given by Eq.(2.15)
\newline
\begin{equation}
\phi(x)=4\hbar^{4}\omega^{2}x(p+1-x)(x+3)(x+2),\quad E=\hbar \omega(p+\frac{8}{3}),
\end{equation}
\newline
The second solution is given by Eq(2.12) or Eq.(2.20) with the corresponding energy spectrum
\newline
\begin{equation}
\phi(x)=4\hbar^{4}\omega^{2}x(p+1-x)(x-3)(x-1),\quad E=\hbar\omega(p-\frac{1}{3}).
\end{equation}
This representation is valid only for p=0.
\newline
We will also treat this systems using the results on supersymmetry. The eigenfunctions for the x part consist of an infinite sequence $\psi_{n}(x)$ starting from $psi_{3}^{0}(x)$ of Eq.(3.37) and a singlet state $\chi(x)$ given by Eq.(3.35) and (3.40)
\begin{equation}
\psi_{n}(x)=N_{n}(a^{\dagger})^{n}e^{\frac{-\omega x^{2}}{2\hbar}}\frac{x(3\hbar +2\omega x^{2})}{(\hbar+2\omega x^{2})}, \chi(x)=C_{0}\frac{e^{\frac{-\omega x^{2}}{2\hbar}}}{\hbar+2\omega x^{2}}
\end{equation}
\begin{equation}
a\chi(x)=0,\quad a^{\dagger}\chi(x)=0\quad .
\end{equation}
The creation and annihilation operator are given by Eq.(3.32) with the following expressions for $W_{1}$, $W_{2}$ and $W_{3}$
\newline
\begin{equation}
W_{1}=\frac{-(-\hbar+2\omega x^{2})(9\hbar^{3}+27\hbar^{2}\omega x^{2}+12\hbar\omega^{2}x^{4}+4\omega^{3}x^{6})}{\hbar x(3\hbar+2\omega x^{2})(3\hbar^{2}+4\omega^{2}x^{4})}\quad ,
\end{equation}
\begin{equation}
W_{2}=\frac{-(\hbar-2\omega x^{2})(3\hbar^{2}+3\hbar\omega x^{2}+2\omega^{2}x^{4})}{\hbar x(3\hbar^{2}+8\hbar\omega x^{2}+4\omega^{2}x^{4})}\quad ,
\end{equation}
\begin{equation}
W_{3}=\frac{-\omega x(-9\hbar^{3}+22\hbar^{2}\omega x^{2}+20\hbar\omega^{2}x^{4}+8\omega^{3}x^{6})}{\hbar(\hbar+2\omega x^{2})(3\hbar^{2}+4\omega^{2}x^{4})}\quad .
\end{equation}
\newline
With the eigenfunctions for the y part of the Hamiltonian and the formula for the energies given by Eq.(3.34) we obtain the two following series of solutions
\newline
\begin{equation}
\psi_{n,k}=\psi_{n}(x)e^{-\frac{\omega y^{2}}{2\hbar}}H_{k}(\sqrt{\frac{\omega}{\hbar}}y),\quad  E=\hbar\omega(n+k+\frac{8}{3}) \quad ,
\end{equation}
\begin{equation}
\phi_{m}=\chi(x)e^{-\frac{\omega y^{2}}{2\hbar}}H_{m}(\sqrt{\frac{\omega}{\hbar}}y), \quad E_{m}=\hbar\omega(m-\frac{1}{3}).
\end{equation}
\newline
\subsection{Case  $ \alpha = 5$, $\beta = -8$, $f(z)=\frac{4z(2z^{2} - 1)(2z^{2}+ 3)}{(2z^{2}+ 1)(4z^{4}+3)}$ and $\epsilon = -1$.}
We have with Eq.(1.4) and (1.5)
\newline
\begin{equation}
V(x,y)=\frac{\omega^{2}}{2}(x^{2}+y^{2})-\frac{192\hbar^{4}\omega^{2}x^{2}}{(4\omega^{2}x^{4}+3\hbar^{2})^{2}}+\frac{16\hbar^{2}\omega^{2}x^{2}}{4\omega^{2}x^{4}+3\hbar^{2}}.
\end{equation}
\newline
From the cubic algebra we obtain three unitary representations. The first unitary representation is given by Eq.(2.32) with the corresponding energy spectrum  Eq.(2.33)
\newline
\begin{equation}
\phi(x)=4\hbar^{4}\omega^{2}x(p+1-x)(x+4)(x+2),\quad E=\hbar \omega(p+3),
\end{equation}
\newline
The second solution is given by Eq(2.38) with the corresponding energy spectrum given by Eq.(2.39)
\newline
\begin{equation}
\phi(x)=4\hbar^{4}\omega^{2}x(p+1-x)(p-3-x)(p-1-x),\quad E=-\hbar\omega(p-1).
\end{equation}
This representation is valid only for p=0,1.
\newline
The third solution is given by Eq(2.30) with the corresponding energy spectrum given by Eq.(2.31)
\newline
\begin{equation}
\phi(x)=4\hbar^{4}\omega^{2}x(p+1-x)(x-3)(x-2),\quad E=\hbar\omega(p-1).
\end{equation}
This representation is valid only for p=0,1.
\newline
We investigate this system using the results on supersymmetry. The eigenfunctions for the x part consist of an infinite sequence $\psi_{n}(x)$ starting from $\psi_{0}^{2}(x)$ given by Eq.(3.22) and doublet states $\chi_{1}(x)$ and $\chi_{2}(x)$ given by Eq.(3.22), (3.21) and (3.26)
\begin{equation}
\psi_{n}(x)=N_{n}(a^{\dagger})^{n}e^{\frac{-\omega x^{2}}{2\hbar}}\frac{(-9\hbar^{3}+18\hbar^{2}\omega x^{2}+12\hbar \omega^{2} x^{4}+8\omega^{3}x^{6})}{(3\hbar^{2}+4\omega^{2}x^{4})},
\end{equation}
\begin{equation}
\chi_{1}(x)=C_{1}e^{\frac{-\omega x^{2}}{2\hbar}}\frac{(\hbar+2\omega x^{2})}{(3\hbar^{2}+4\omega^{2}x^{4})},\quad \chi_{2}(x)=C_{2}e^{\frac{-\omega x^{2}}{2\hbar}}\frac{x(3\hbar+2\omega x^{2})}{(3\hbar^{2}+4\omega^{2}x^{4})}.
\end{equation}
\begin{equation}
a\chi_{1}(x)=0,\quad a^{\dagger}\chi_{1}(x)=\chi_{2}(x),\quad a^{\dagger}\chi_{2}(x)=0\quad .
\end{equation}
The creation and annihilation operators are given by Eq.(3.6) with $W_{1}$, $W_{2}$ and $W_{3}$ as in Eq.(4.5), Eq.(4.6) and Eq.(4.7).
\newline
With the eigenfunctions in the y part and the formula for the energies given by Eq.(3.20) we obtain the three following kinds of solutions
\newline
\begin{equation}
\psi_{n,k}=\phi_{n}(x)e^{-\frac{\omega y^{2}}{2\hbar}}H_{k}(\sqrt{\frac{\omega}{\hbar}}y),\quad E=\hbar\omega(n+k+3) \quad ,
\end{equation}
\begin{equation}
\phi_{m_{1}}=\chi_{1}(x)e^{-\frac{\omega y^{2}}{2\hbar}}H_{m_{1}}(\sqrt{\frac{\omega}{\hbar}}y), \quad E_{m_{1}}=\hbar\omega(m_{1}-1)\quad ,
\end{equation}
\begin{equation}
\phi_{m_{2}}=\chi_{2}(x)e^{-\frac{\omega y^{2}}{2\hbar}}H_{m_{2}}(\sqrt{\frac{\omega}{\hbar}}y),\quad E_{m_{2}}=\hbar\omega m_{2}.
\end{equation}
\subsection{Case $ \alpha = 0$, $\beta = -\frac{2}{9}$, $f(z)=-\frac{2}{3}z$ and $\epsilon = 1$.}
We have
\newline
\begin{equation}
V(x,y)=\frac{\omega^{2}}{2}(\frac{1}{9}x^{2}+y^{2})
\end{equation}
\newline
From the cubic algebra we get three cases with unitary representations and using Eq.(2.12) to Eq.(2.17) we have 
\newline
\begin{equation}
\phi(x)=4\hbar^{4}\omega^{2}x(p+1-x)(x+\frac{1}{3})(x+\frac{2}{3}),\quad E=\hbar \omega(p+\frac{4}{3}),
\end{equation}
\begin{equation}
\phi(x)=4\hbar^{4}\omega^{2}x(p+1-x)(x-\frac{1}{3})(x+\frac{1}{3}),\quad E=\hbar \omega(p+1),
\end{equation}
\begin{equation}
\phi(x)=4\hbar^{4}\omega^{2}x(p+1-x)(x-\frac{2}{3})(x-\frac{1}{3}),\quad E=\hbar \omega(p+\frac{2}{3}).
\end{equation}
\newline
We apply the results coming from supersymmetry. we get the following known eigenfunctions from Eq.(3.35), Eq.(3.37) and E.(3.36) respectively and the corresponding energy with the Eq.(3.34)
\begin{equation}
\psi_{n_{1},k_{1}}=N_{n_{1}k_{1}}(a^{\dagger})^{n_{1}}e^{\frac{-\omega x^{2}}{6\hbar}}(-3\hbar +2\omega x^{2})e^{-\frac{\omega y^{2}}{2\hbar}}H_{k_{1}}(\sqrt{\frac{\omega}{\hbar}}y),
\end{equation}
\begin{equation}
E_{1}=\hbar\omega(n_{1}+k_{1}+\frac{4}{3}),
\end{equation}
\begin{equation}
\psi_{n_{2},k_{2}}=N_{n_{2}k_{2}}(a^{\dagger})^{n_{2}}e^{\frac{-\omega x^{2}}{6\hbar}}xe^{-\frac{\omega y^{2}}{2\hbar}}H_{k_{2}}(\sqrt{\frac{\omega}{\hbar}}y),
\end{equation}
\begin{equation}
E_{2}=\hbar\omega(n_{2}+k_{2}+1).
\end{equation}
\begin{equation}
\psi_{n_{3},k_{3}}=N_{n_{3}k_{3}}(a^{\dagger})^{n_{3}}e^{\frac{-\omega x^{2}}{6\hbar}}e^{-\frac{\omega y^{2}}{2\hbar}}H_{k_{3}}(\sqrt{\frac{\omega}{\hbar}}y),
\end{equation}
\begin{equation}
E_{2}=\hbar\omega(n_{2}+k_{2}+\frac{2}{3}),
\end{equation}
\newline
The creation and annihilation operator are given by Eq.(3.32) with the following expression for $W_{1}$, $W_{2}$ and $W_{3}$
\newline
\begin{equation}
W_{1}=\frac{1}{x}+\frac{\omega x}{3\hbar},\quad W_{2}=-\frac{1}{x}+\frac{\omega x}{3\hbar},\quad W_{3}=-\frac{\omega x}{3\hbar}.
\end{equation}
\subsection{$ \alpha = -1, \beta = -\frac{32}{9}$, $f(z)= -\frac{2z}{3} - \frac{2z^{2}- 3}{z(2z^{2}+ 3)}$ and $\epsilon = 1$.}
We have
\newline
\begin{equation}
V(x,y)=\frac{\omega^{2}}{2}(\frac{1}{9}x^{2}+y^{2})-\frac{24\hbar^{3}\omega}{(2\omega x^{2}+3\hbar)^{2}}+\frac{4\hbar^{2}\omega}{(2\omega x^{2}+3\hbar)} \quad .
\end{equation}
\newline
\newline
From the cubic algebra we get the three cases with unitary representations from Eq.(2.12) to Eq.(2.17)
\newline
\begin{equation}
\phi(x)=4\hbar^{4}\omega^{2}x(p+1-x)(x+\frac{1}{3})(x+\frac{5}{3}),\quad E=\hbar \omega(p+\frac{5}{3}),
\end{equation}
\begin{equation}
\phi(x)=4\hbar^{4}\omega^{2}x(p+1-x)(x-\frac{1}{3})(x+\frac{4}{3}),\quad E=\hbar \omega(p+\frac{4}{3}),
\end{equation}
\begin{equation}
\phi(x)=4\hbar^{4}\omega^{2}x(p+1-x)(x-\frac{5}{3})(x-\frac{4}{3}),\quad E=\hbar \omega(p).
\end{equation}
\newline
From the supersymmery we obtain the following eigenfunctions from Eq.(3.35), Eq.(3.36) and E.(3.37) respectively and the energy with the Eq.(3.34)
\begin{equation}
\psi_{n_{1},k_{1}}=N_{n_{1}k_{1}}(a^{\dagger})^{n_{1}}e^{\frac{-\omega x^{2}}{6\hbar}}x\frac{(-45\hbar^{2}+4\omega^{2}x^{4})}{(3\hbar + 2\omega x^{2})}e^{-\frac{\omega y^{2}}{2\hbar}}H_{k_{1}}(\sqrt{\frac{\omega}{\hbar}}y),
\end{equation}
\begin{equation}
E_{1}=\hbar\omega(n_{1}+k_{1}+\frac{5}{3}),
\end{equation}
\begin{equation}
\psi_{n_{2},k_{2}}=N_{n_{2}k_{2}}(a^{\dagger})^{n_{2}}e^{\frac{-\omega x^{2}}{6\hbar}}\frac{(9\hbar^{2}-12\hbar\omega x^{2}-4\omega^{2}x^{4})}{(3\hbar +2\omega x^{2})}e^{-\frac{\omega y^{2}}{2\hbar}}H_{k_{2}}(\sqrt{\frac{\omega}{\hbar}}y),
\end{equation}
\begin{equation}
E_{2}=\hbar\omega(n_{2}+k_{2}+\frac{4}{3}).
\end{equation}
\begin{equation}
\psi_{n_{3},k_{3}}=N_{n_{3}k_{3}}(a^{\dagger})^{n_{3}}\frac{e^{\frac{-\omega x^{2}}{6\hbar}}}{(3\hbar +2\omega x^{2})}e^{-\frac{\omega y^{2}}{2\hbar}}H_{k_{3}}(\sqrt{\frac{\omega}{\hbar}}y).
\end{equation}
\begin{equation}
E_{3}=\hbar\omega(n_{3}+k_{3}),
\end{equation}
The creation and annihilation operators are given by Eq.(3.32) with the following expressions for $W_{1}$, $W_{2}$ and $W_{3}$
\newline
\begin{equation}
W_{1}=\frac{-27\hbar^{3}+27\hbar^{2}\omega x^{2}+48\hbar\omega^{2}x^{4}+4\omega^{3}x^{6}}{27\hbar^{3}x-36\hbar^{2}\omega x^{3}-12\hbar\omega^{2}x^{5}},
\end{equation}
\begin{equation}
W_{2}=\frac{351\hbar^{3}\omega x+126\hbar^{2}\omega^{2}x^{3}+12\hbar\omega^{3}x^{5}-8\omega^{4}x^{7}}{81\hbar^{4}-54\hbar^{3}\omega x^{2}-106\hbar^{2}\omega^{2}x^{4}-24\hbar\omega^{3}x^{6}},
\end{equation}
\begin{equation}
W_{3}=\frac{-9\hbar^{2}-3\hbar\omega x^{2}+2\omega^{2}x^{4}}{9\hbar^{2}x+6\hbar\omega x^{3}}.
\end{equation}
\subsection{Case  $ \alpha = 0$, $\beta = -2$, $f(z)=-2z-\Psi(z)$ and $\epsilon =1$.}
We have 
\newline
\begin{equation}
\Psi(z)=\frac{\psi'(z)}{\psi(z)},\quad \psi(z)=1 - t E_{c}(z) \quad .
\end{equation}
$E_{c}(z)$ is the complementary error function and is given by
\newline
\begin{equation}
E_{c}(z)=\frac{2}{\sqrt{\pi}}\int_{z}^{\infty}e^{-t^{2}}dt
\end{equation}
We have with Eq.(1.4) and (1.5) 
\newline
\begin{equation}
V(x,y)=\frac{\omega^{2}}{2}(x^{2}+y^{2})-\frac{2}{3}\hbar\omega+\frac{4e^{-\frac{2\omega x^{2}}{h}}t\hbar\omega}{\pi(1-tE_{c}(\sqrt{\frac{\omega}{\hbar}}x))^{2}}+\frac{4e^{-\frac{\omega x^{2}}{h}}+\omega\sqrt{\hbar\omega}x}{\pi(1-tE_{c}(\sqrt{\frac{\omega}{\hbar}}x))}.
\end{equation}
\newline
The cubic algebra provides two unitary representations. The first unitary representation is given by the Eq.(2.12) with the corresponding energy given by Eq.(2.13). The Eq.(2.14) and (2.15) give the same unitary representation and energy spectrum and we have
\newline
\begin{equation}
\phi(x)=4\hbar^{4}\omega^{2}x(p+1-x)x(x+1),\quad E=\hbar \omega(p+\frac{4}{3}),
\end{equation}
The second solution is given by Eq(2.16) with the corresponding energy spectrum given by Eq.(2.17)
\newline
\begin{equation}
\phi(x)=4\hbar^{4}\omega^{2}x(p+1-x)(x-1)^{2},\quad E=\hbar\omega(p+\frac{1}{3}).
\end{equation}
This representation is valid for p=0.
\newline
We also use supersymmetry to treat this system. The eigenfunctions for the x part consist of an infinite sequence $\psi_{n}(x)$ starting from $\psi_{3}^{0}$ given by Eq.(3.37) and a singlet state $\chi(x)$ given by Eq.(3.36) and (3.39)
\begin{equation}
\psi_{n}(x)=N_{n}(a^{\dagger})^{n}e^{\frac{-3\omega x^{2}}{2\hbar}}\frac{(-t\sqrt{\hbar\omega}-e^{\frac{\omega x^{2}}{\hbar}}\sqrt{\pi}\omega x+e^{\frac{\omega x^{2}}{\hbar}}\sqrt{\pi}t\omega x E_{c}(\sqrt{\frac{\omega}{\hbar}}x))}{\omega(-1+tE_{c}(\sqrt{\frac{\omega}{\hbar}}x))},
\end{equation}
\begin{equation}
\chi(x)=C_{0}\frac{e^{\frac{-\omega x^{2}}{2\hbar}}}{\sqrt{\pi}\hbar(1-tE_{c}(\sqrt{\frac{\omega}{\hbar}}x))}\quad ,
\end{equation}
\begin{equation}
a\chi(x)=0,\quad a^{\dagger}\chi(x)=0\quad .
\end{equation}
The creation and annihilation operators are given by Eq.(3.32) with the following expressions for $W_{1}$, $W_{2}$ and $W_{3}$
\newline
\begin{equation}
W_{1}=\frac{(-t\omega x-e^{\frac{\omega x^{2}}{\hbar}}\sqrt{\pi}\sqrt{\frac{\omega}{\hbar}}(\hbar+\omega x^{2})+e^{\frac{\omega x^{2}}{\hbar}}\sqrt{\pi}t\sqrt{\frac{\omega}{\hbar}}(\hbar+\omega x^{2})E_{c}(\sqrt{\frac{\omega}{\hbar}}x))}{\hbar(-t-e^{\frac{\omega x^{2}}{\hbar}}\sqrt{\pi}\sqrt{\frac{\omega}{\hbar}}x+e^{\frac{\omega x^{2}}{\hbar}}\sqrt{\pi}+\sqrt{\frac{\omega}{\hbar}}xE_{c}(\sqrt{\frac{\omega}{\hbar}}x))}    \quad ,
\end{equation}
\begin{equation}
W_{2}=\frac{A}{B}   \quad ,
\end{equation}
\begin{equation}
A=e^{-\frac{\omega x^{2}}{\hbar}}(2t^{2}\sqrt{\omega\hbar}+3e^{\frac{\omega x^{2}}{\hbar}}\sqrt{\pi}t\omega x-e^{\frac{2\omega x^{2}}{\hbar}}\pi\sqrt{\frac{\omega}{\hbar}}(h - \omega x^{2})+
\end{equation}
\[(-3e^{\frac{\omega x^{2}}{h}}\sqrt{\pi}t^{2}\omega x+2e^{\frac{2\omega x^{2}}{\hbar}}\pi t\sqrt{\frac{\omega}{\hbar}}(\hbar - \omega x^{2}))E_{c}(\sqrt{\frac{\omega}{\hbar}}x)                   \]
\[ -e^{\frac{2\omega x^{2}}{\hbar}}\pi t^{2}\sqrt{\frac{\omega}{\hbar}}(\hbar - \omega x^{2})(E_{c}(\sqrt{\frac{\omega}{\hbar}}x))^{2})  ,                   \]
\begin{equation}
B=\hbar\sqrt{\pi}(-1+tE_{c}(\sqrt{\frac{\omega}{\hbar}}x))(-t-e^{\frac{\omega x^{2}}{\hbar}}\sqrt{\pi}\sqrt{\frac{\omega}{\hbar}}x+e^{\frac{\omega x^{2}}{\hbar}}\sqrt{\pi}t\sqrt{\frac{\omega}{\hbar}}xE_{c}(\sqrt{\frac{\omega}{\hbar}}x)),
\end{equation}
\begin{equation}
W_{3}=\frac{\omega x}{\hbar}+\frac{2e^{-\frac{\omega x^{2}}{\hbar}}+\sqrt{\frac{\omega}{\hbar}}}{\sqrt{\pi}(1-tE_{c}(\sqrt{\frac{\omega}{\hbar}}x))}    \quad .  
\end{equation}
\newline
With the eigenfunctions for the y part of the Hamiltonian and the formula for the energies given by Eq.(3.34) we obtain the two following families of solutions
\newline
\begin{equation}
\psi_{n,k}=\psi_{n}(x)e^{-\frac{\omega y^{2}}{2\hbar}}H_{k}(\sqrt{\frac{\omega}{\hbar}}y),\quad  E=\hbar\omega(n+k+\frac{4}{3}) \quad ,
\end{equation}
\begin{equation}
\phi_{m}=\chi(x)e^{-\frac{\omega y^{2}}{2\hbar}}H_{m}(\sqrt{\frac{\omega}{\hbar}}y), \quad E_{m}=\hbar\omega(m+\frac{1}{3}).
\end{equation}
\newline
\subsection{Case  $ \alpha = 0$, $\beta = -2$, $f(z)=-2z-\Psi(z)$ and $\epsilon =-1$.}
This case give the harmonic oscillator. The zero mode is given by Eq.(3.22) is the well known ground state of the harmonic oscillator. There is other special solutions in terms of the complementary error function exist.
\newline
Many special solutions of the fourth Painlev\'e equation will give us singular Hamiltonians. These potentials can be regularized in several manners [19,56,57,58]. 
\section{Conclusion}
The main results of this article are that we have constructed the cubic algebra, Fock type presentations and the corresponding energy spectrum for the potential given by Eq.(1.4), (1.5). Other superintegrable potentials written in term of Painlev\'e transcendents are known [16] namely:
\newline
\begin{equation}
V_{1}=\hbar^{2}(\omega_{1}^{2}P_{I}(\omega_{1}x)+\omega_{2}^{2}P_{I}(\omega_{2}y))
\end{equation}
\begin{equation}
V_{2}=ay+\hbar^{2}\omega ^{2}P_{I}(\omega x)
\end{equation}
\begin{equation}
V_{3}=bx+ay+(2\hbar b)^{\frac{2}{3}}P_{II}^{2}((\frac{2b}{\hbar^{2}})^{\frac{1}{3}}x,0)
\end{equation}
\begin{equation}
V_{4}=ay+(2\hbar^{2} b^{2})^{\frac{1}{3}}(P'_{II}((\frac{-4b}{\hbar^{2}})^{\frac{1}{3}}x,k)+P_{II}^{2}((\frac{-4b}{\hbar^{2}})^{\frac{1}{3}}x)                    ,k).
\end{equation}
\newline
These potentials together with that in Eq.(1.4) and Eq.(1.5) were also obtained as one dimensional potentials in the context of higher and conditional symmetries by W.I.Fushchych and A.G.Nikitin [46]. For these four superintegrable potentials the simplest underlying structure of the type (2.5) is actually a finite dimensional Lie algebra that does not allow us to find the energy spectrum. Let us present these algebras:
\newline
For $V_{1}$ we have:
\newline
\begin{equation}
[A,B]\equiv C=-i\hbar^{5}\omega_{1}^{5}\omega_{2}^{5},\quad [A,C]=0,\quad [B,C]=0\quad .
\end{equation}
\newline
For $V_{2}$ we have:
\newline
\begin{equation}
[A,B]\equiv C=-i\hbar^{5}\omega^{5},\quad [A,C]=0,\quad [B,C]=0\quad .
\end{equation}
\newline
For $V_{3}$ and $V_{4}$ we have:
\newline
\begin{equation}
[A,B]\equiv C=4abi\hbar(H+\frac{1}{2}A),\quad [A,C]=0,\quad [B,C]=8a^{2}b^{2}\hbar^{2}(H+\frac{1}{2}A)\quad .
\end{equation} 
\newline
These algebras coincide with the classical Poisson algebras presented earlier [17]. In these four cases we have a triplet of commuting operators.
The x part of the potential $V_{4}$ was also obtained in the context of supersymmetric quantum mechanics [55]. The methods of this article are not directly applicable in that case, but it may be possible to generalize them.
\newline
The question of how these aspects of SUSYQM, shape invariance and superintegrability are
related is interesting and will require more study. Supersymmetry could also be a tool for the classification of superintegrable potentials. Higher order supersymmetry could be a suitable approach to treat these potentials. The search for superintegrable systems with higher order integrals of motion is thus closely related to the subject of polynomial algebras and higher order supersymmetric quantum mechanics. 
\newline
The search for a grand unifying theory in particle physics is an important problem of comtemporary physics. One model that is envisaged as a candidate is string theory. The x part of the potential given by Eq.(1.1) appears also in the context of string theory [59] where supersymmetry is used as a method for constructing exact solutions. A more recent article [60] discusses how supersymmetric quantum mechanics can be used to construct solutions in string theory.
\newline
\newline
\textbf{Acknowledgments} The research of I.M. is supported by a doctoral
research scholarship from FQRNT of Quebec. The author thanks P.Winternitz for very helpful comments and discussions.
This article was written in part while he was visiting the Universita di Roma Tre. He thanks D.Levi for his hospitality.
He thanks the University of Montreal and the Ministry of Education of Quebec for the PBCSE travel awards.

\section{\textbf{References}}
1. V.Fock, Z.Phys. 98, 145-154 (1935).
\newline
2. V.Bargmann, Z.Phys. 99, 576-582 (1936).
\newline
3. J.M.Jauch and E.L.Hill, Phys.Rev. 57, 641-645 (1940).
\newline
4. M.Moshinsky and Yu.F.Smirnov, The Harmonic Oscillator In Modern
Physics, (Harwood, Amsterdam, 1966).
\newline
5. J.Fris, V.Mandrosov, Ya.A.Smorodinsky, M.Uhlir and P.Winternitz, Phys.Lett. 16, 354-356 (1965).
\newline
6. P.Winternitz, Ya.A.Smorodinsky, M.Uhlir and I.Fris, Yad.Fiz. 4,
625-635 (1966). (English translation in Sov. J.Nucl.Phys. 4,
444-450 (1967)).
\newline
7. A.Makarov, Kh. Valiev, Ya.A.Smorodinsky and P.Winternitz, Nuovo Cim. A52, 1061-1084 (1967).
\newline
8. N.W.Evans, Phys.Rev. A41, 5666-5676 (1990), J.Math.Phys. 32,
3369-3375 (1991).
\newline
9. E.G.Kalnins, J.M.Kress, W.Miller Jr and P.Winternitz,
J.Math.Phys. 44(12) 5811-5848 (2003).
\newline
10. E.G.Kalnins, W.Miller Jr and G.S.Pogosyan, J.Math.Phys. A34,
4705-4720 (2001).
\newline
11. E.G.Kalnins, J.M.Kress and W.Miller Jr, J.Math.Phys. 46,
053509 (2005), 46, 053510 (2005), 46, 103507 (2005), 47, 043514
(2006), 47, 043514 (2006).
\newline
12. E.G.Kalnins, W.Miller Jr and G.S.Pogosyan, J.Math.Phys. 47,
033502.1-30 (2006), 48, 023503.1-20 (2007).
\newline
13. J.Drach, C.R.Acad.Sci.III, 200, 22 (1935), 200, 599 (1935).
\newline
14. J.Hietarinta, Phys.Lett.A, 246, 1, 97 (1998).
\newline
15. S.Gravel and P.Winternitz, J.Math.Phys. 43(12), 5902 (2002).
\newline
16. S.Gravel, J.Math.Phys. 45(3), 1003-1019 (2004).
\newline
17. I.Marquette and P.Winternitz, J.Math.Phys. 48(1) 012902
(2007).
\newline
18. I.Marquette and P.Winternitz, J. Phys. A: Math. Theor. 41, 304031 (2008).
\newline
19. I.Marquette, J. Math. Phys. 50, 012101 (2009).
\newline
20. P.Letourneau and L.Vinet, Ann.Phys. 243, 144 (1995).
\newline
21. Ya. I Granovskii, A.S. Zhedanov  and I.M. Lutzenko , J.
Phys. A24, 3887-3894 (1991).
\newline
22. D.Bonatsos, C.Daskaloyannis and K.Kokkotas, Phys.Rev. A48,
R3407-R3410 (1993).
\newline
23. D.Bonatsos, C.Daskaloyannis and K.Kokkotas, Phys. Rev.A 50, 3700-3709 (1994).
\newline
24. P.Letourneau and L.Vinet, Ann.Phys. 243, 144 (1995).
\newline
25. C.Daskaloyannis, J.Math.Phys. 42, 1100 (2001).
\newline
26. C.Daskaloyannis, Generalized deformed oscillator and nonlinear
algebras, J.Phys.A: Math.Gen 24, L789-L794 (1991).
\newline
27. E.G.Kalnins, W.Miller and S.Post, SIGMA 4, 008 (2008). 
\newline
28. C.Quesne, SIGMA 3, 067 (2007).
\newline
29. V.Sunilkumar, arXiv:math-ph/0203047 (2002).
\newline
30. V.Sunilkumar, B.A.Bambah and R.Jagannathan, J.Phys. A:Math.Gen.34, 8583 (2001).
\newline
31. V.Sunilkumar, B.A.Bambah and R.Jagannathan, mod. Phys. Lett. A17, 1559 (2002).
\newline
32.  E.Witten, Nucl.Phys. B188, 513 (1981).
\newline
33. G.Darboux, C.R.Acad.Sci. Paris, 94, 1459 (1882)
\newline
34. T.F.Moutard, C.R.Acad.Sci. Paris, 80, 729 (1875), J.de L'ï¿½cole Politech., 45, 1 (1879).
\newline
35. E.Schrodinger, Proc.Roy. Irish Acad., 46A, 9 (1940), 47A, 53 (1941).
\newline
36. L.Infeld and T.E.Hull, Rev.Mod.Phys., 23, 21 (1951).
\newline
37. G.Junker, Supersymmetric Methods in Quantum and Statistical Physics, Springer, New York, (1995).
\newline
38. M.S.Plyushchay, Annals Phys. (N.Y.) 245, 339 (1996)
\newline
39. M.Plyushchay, Int.J.Mod.Phys. A15, 3679 (2000), 
\newline 
40. F.Correa and M.Plyushchay, Annals Phys.322, 2493 (2007).
\newline
41. L.Gendenshtein, JETP Lett., 38, 356 (1983).
\newline
42. E.L.Ince, Ordinary Differential Equations (Dover, New York,
1944).
\newline
43. M.J.Ablowitz and P.A.Clarkson, Solitons, Nonlinear Evolution
Equations and Inverse Scattering, (Cambridge Univ.Press, 1991).
\newline
44. A.P.Veselov and A.Shabat,Funkt.Analiz.Prilozh 27(2), 1-21 (1993).
\newline
45. A.P.Veselov, J.Phys. A34, 3511 (2001).
\newline
46. V.Spiridonov, Phys.Rev.A52, 3 (1995).
\newline
47. W.I. Fushchych and A.G. Nikitin, J.Math.Phys. V38 N11, 5944 (1997).
\newline
48. V. Gromak, I Laine and S.Shimomura, Painlev\'e Differential Equations
in the Complex Plane, Walter de Gruyter (2002).
\newline
49. A.Andrianov, M.Ioffe and V.P.Spiridonov, Phys.Lett. A174, 273 (1993). 
\newline
50. A.Andrianov, F.Cannata, J.P.Dedonder and M.Ioffe, Int.Mod.Phys.A10, 2683-2702 (1995).
\newline
51. A.Andrianov, F.Cannata, M.Ioffe and D.Nishnianidze, Phys.Lett.A, 266,341-349 (2000).
\newline
52. M.V.Ioffe and D.N.Mishnianidze, PhysLett.A327,425 (2004)
\newline
53. A.A.Andrianov and A.V.Sokolov, Jour.Math.Sci, 143, 1, 2702 (2007).
\newline
54. J. Mateo and J. Negro, J.Phys. A:Math. Theor. 41 045204 (2008).
\newline
55. F.Cannata, M.Ioffe, G.Junker and D.Nishnianidze, J.Phys.A.Math.Gen. 32, 3583 (1999)
\newline
56. I.F.Marquez, J.Negro and L.M.Nieto, J.Phys. A:Math. Gen. 31, 4115 (1998).
\newline
57. A.Das and S.A.Pernice, Nucl.Phys.B 561, 357 (1999).
\newline
58. M.Znojil,Nucl.Phys.B662,554 (2003), M.Znojil, Phys.Lett. A 259, 220 (1999).
\newline
59. A.V.Yurov and V.A.Yurov, Phys.Rev.D72, 026003 (2005).
\newline
60. M.R.Setare, J.Sadeghi and A.R.Amani, Phys.Lett B 660, 299 (2008).
\newline

\end{flushleft}
\end{document}